\documentclass{jetpl}
\twocolumn

\lat

\title{Nucleon form factors for the elastic electron-deuteron scattering at high momentum transfer}

\rtitle{Nucleon form factors for the elastic electron-deuteron scattering at high momentum transfer}

\sodtitle{Nucleon form factors for the elastic electron-deuteron scattering at high momentum transfer}

\author{A.\,V.\,Bekzhanov$^{+*}$\/\thanks{e-mail: bekzhanov@jinr.ru}, S.\,G.\,Bondarenko$^+$, V.\,V.\,Burov$^+$}

\rauthor{A.\,V.\,Bekzhanov, S.\,G.\,Bondarenko, V.\,V.\,Burov}

\sodauthor{Bekzhanov, Bondarenko, Burov}

\address{$^+$Joint Institute for Nuclear Research,
141980 Dubna, Moscow region, Russia\\~\\
$^*$FEFU - Far Eastern Federal University, 8 Suhanova, 690950 Vladivostok, Russia}

\dates{}{*}

\abstract{
The reaction of the elastic electron-deuteron scattering at high momentum
transfer is investigated within the Bethe-Salpeter approach.
The relativistic covariant Graz II separable kernel
of nucleon-nucleon interactions is used to analyze the deuteron
structure functions, form factors and tensor of polarization components.
The modern data for the electromagmetic nucleons structure from
the double polarization experiments as well as some other models
of the nucleon form factors are considered.
}


\begin{document}

\maketitle

\section{Introduction}
The deuteron being the simplest two-nucleon bound system is a powerful
instrument to study strong interactions. The reaction of elastic
electron-deuteron scattering provides information not only on NN
interaction but also on the electromagnetic structure of nucleons.
Such investigations at high energies are of great interest nowadays,
especially in the context of future experiments at being upgraded JLab
facilities.

It is necessary  to note that in order to describe the elastic form factors
of the deuteron at high momentum transfer ($Q^2=-q^2 > 2$ (GeV/c)$^2$)
the relativistic properties of the strong interactions should be taken
into account. Here, properties of core nuclear forces play a very important
role. From the physical point of view, elastic electron-deuteron
scattering at transfer
momentum up to 6 (GeV/c)$^2$ is an amazing phenomenon taking into account that
binding energy of the deuteron is very small (2.2 MeV). So the subject
of investigation has a great significance for the nuclear and particle physics.

Some approaches, based on the Bethe-Salpeter (BS)
equation~\cite{Salpeter:1951sz}, satisfy this condition, among them are the
light-front dynamics~\cite{karmanov}, the equal-time
equation~\cite{Pascalutsa:1997vk}, BS approach with separable
interaction~\cite{Bondarenko:2002zz} and so on.
In the last approach, one has to solve the system of linear integral
equations for both the NN scattered states and the bound state --
the deuteron. In order to find a solution of  a system of
integral equations, it is a good idea to use a separable
ansatz~\cite{Bondarenko:2002zz} for the interaction kernel in the
BS equation. Then, one can transform the integral equations
into a system of algebraic linear ones which can be solved.
Parameters of the interaction kernel are extracted from an analysis
of phase shifts for respective partial-wave states and low-energy
parameters as well as deuteron properties (bound state
energy, magnetic moment, elastic form factors etc.).
In the Refs.~\cite{Bondarenko:2008mm} and~\cite{Bondarenko:2010rv}
the latter approach was developed and applied to the reaction
of the deuteron electrodisintegration.

The electromagnetic (EM) structure of nucleons at high momentum transfer
is another topic of interest. In the paper, four models for the nucleon form
factors are used. First of them -- the dipole fit (DFF)~\cite{Pietschmann:1969mj} --
was widely used. The main feature of this model is that the ratio of electric
$G^{\rm E}_{\rm p}$ and magnetic $G^{\rm M}_{\rm p}$ proton form factors
is constant. Another one -- the relativistic harmonic oscillator model (RHOM)~\cite{Burov:1993ia} --
is the quark model with a relativistic harmonic oscillator potential.

However, recently there was an intensive discussion that the ratio
obtained by the Rosenbluth separation technique differs from the
one obtained by the recoil polarization method~\cite{Gayou:2001qd,Jones:1999rz}.
To describe the results of the latter method, it is necessary to use a
certain parametrization of the ratio as some linear function of the transfer
momentum squared. The model with described ratio for the electric proton form factor
and the Galster parametrization~\cite{Galster:1971kv} for the
neutron electric form factor -- modified dipole fit (MDFF1) - is also considered
(see also,~\cite{Egiyan:2007qj} and~\cite{Bondarenko:2010rv}).

Recently the Unitary and Analytic (U\&A) approach has been used to develop new
nine-resonance model~\cite{Adamuscin:2013zoa}. This model which includes
new experimental data on the nucleon EM form factors as well as a new method
of introducing the asymptotic behavior for the EM form factors also used
in calculations.

In contrast to the Ref.~\cite{Rupp:1989sg}, the influence of the new
parametrization of proton electric form factor is investigated, and
in the development of our previous papers~\cite{Bondarenko:2000yg}
and~\cite{Bekzhanov:2013yoa}, the deuteron form factors are calculated
at high energies where the analytic structure of the vertex functions
should be taken into account.

Also we have considered high-energy dynamics of the
poles contributions which arise from the analytic structure of the separable
kernel.

The paper is organized as follows: in Sec.2 we describe the models of the
nucleon form factors and consider analytic structure of the deuteron
EM current in the Sec.3. The obtained results are discussed in Sec.4. In Sec.5
the conclusion is given.

\section{Models of the nucleon form factors}
We calculate elastic electron-deuteron scattering in the relativistic impulse
approximation within the BS approach with the covariant Graz II (rank III)
kernel of the NN interaction~\cite{Mathelitsch:1981mr} and~\cite{Rupp:1989sg}.

Details of the calculations of the deuteron structure functions
$A(q^2)$, $B(q^2)$, charge $F_{\rm C}(q^2)$, quadrupole $F_{\rm Q}(q^2)$ and
magnetic $F_{\rm M}(q^2)$ form factors and tensor polarization
components $T_{20}$, $T_{21}$ of the final deuteron can be found
in~\cite{Bondarenko:2000yg}.

We use four models of the electromagnetic nucleon form factors
(see also,~\cite{Pietschmann:1969mj},~\cite{Egiyan:2007qj},~\cite{Adamuscin:2013zoa} and
~\cite{Burov:1993ia}):
\begin{itemize}
\item the original dipole fit for the proton and neutron form factors (DFF) is
\begin{eqnarray}
&&F_d = (1+Q^2/0.71)^{-2},\nonumber\\
&&G_{\rm E}^{\rm p}=F_d,\hskip 15mm
G_{\rm E}^{\rm n}=0,\nonumber\\
&&G_{\rm M}^{\rm p} = \mu_{\rm p}G_{\rm E}^{\rm p},\hskip 10mm
G_{\rm M}^{\rm n} = \mu_{\rm n}G_{\rm E}^{\rm p};
\label{dff}
\end{eqnarray}
\item the modified dipole fit 1 (MDFF1) is
\begin{eqnarray}
&&G_{\rm E}^{\rm p}=(1-0.13(Q^2-0.04))F_d,\nonumber\\
&&G_{\rm E}^{\rm n}=-\frac{\mu_n\tau}{1+5.6\tau}F_d,\nonumber\\
&&G_{\rm M}^{\rm p} = \mu_{\rm p}F_d,\nonumber\\
&&G_{\rm M}^{\rm n} = \mu_{\rm n}F_d.
\label{mdff1}
\end{eqnarray}
In MDFF1 we take into account the latest JLab data~\cite{Gayou:2001qd}
for the proton electric form factor by the following ratio
$\mu_p G_{\rm E}^{\rm p}/G_{\rm M}^{\rm p}=1-0.13(Q^2-0.04)$,
while for the neutron electric form factor we use the
Galster~\cite{Galster:1971kv} parametrization.

\item the proposed nine-resonance U\&A model
of the nucleon has 12 free parameters. Their values were obtained from the
analysis of the existent experimental data and additionally new one measured
recently in Mainz. All details and formulas can be found in
Ref.~\cite{Adamuscin:2013zoa}.

\item the relativistic harmonic oscillator is
\begin{eqnarray}
&&I^{(3)}=\frac{1}{(1+Q^2/2m^2)^2}\nonumber\\
&&\times\exp{\frac{1}{2\cdot0.42}\frac{-Q^2}{1+Q^2/2m^2}},\nonumber\\
&&G_{\rm E}^{\rm p}=I^{(3)},\nonumber\\
&&G_{\rm E}^{\rm n}=Q^2/2m^2 I^{(3)},\nonumber\\
&&\frac{G_{\rm M}^{\rm p}}{\mu_{\rm p}}=\frac{G_{\rm M}^{\rm n}}{\mu_{\rm n}}=I^{(3)}.
\label{rhom}
\end{eqnarray}
The relativistic harmonic oscillator model is based on the quark model with the
relativistic oscillator potential. All the FFs calculated in this model
have the correct asymptotic behavior. The only free parameter in the model is
the oscillator parameter which was found from fitting of the experimental data.
\end{itemize}

Above $\mu_{\rm p}=2.7928$ and $\mu_{\rm n}=-1.9130$ are the magnetic moments
of the nucleons, $Q^2=-q^2>0$ is the transfer momentum squared,
$\tau=Q^2/4m^2$, $m$ is the nucleon mass, and all dimensional parameters
are in (GeV/c)$^2$.

\section{Analytic structure}
After the partial-wave decomposition the matrix element of the deuteron
current has the following form
\begin{eqnarray}
\langle D^{\prime}{\cal M}^{\prime} | j_{\mu} | D {\cal M} \rangle\nonumber
={\cal I}_{1\;\mu}^{{\cal M}^{\prime}{\cal M}}(q^2)\;F_1^{\rm (S)}(q^2)+\\
+{\cal I}_{2\;\mu}^{{\cal M}^{\prime}{\cal M}}(q^2)\;F_2^{\rm (S)}(q^2),\label{dcur-part}\\
{\cal I}_{1,2\;\mu}^{{\cal M}^{\prime}{\cal M}}(q^2)=i\int dp_0\;|\textbf{p}|^2\;d|\textbf{p}|\;d(\cos{\theta})\nonumber\\
\times \sum_{L\prime,L=0,2}\;
\phi_{L\prime}(p_0^{\prime},|\textbf{p}^{\prime}|)\phi_{L}(p_0,|\textbf{p}|)\nonumber\\
\times I^{L^{\prime},L}_{1,2\;{\cal M^{\prime}}{\cal M}\;\mu}(p_0,|\textbf{p}|,\cos{\theta},q^2),\nonumber
\end{eqnarray}
where the function $I^{L^{\prime},L}_{1,2\;{\cal M^{\prime}}{\cal M}\;\mu}
(p_0,|\textbf{p}|,\cos{\theta},q^2)$ is the result of the trace calculations.
The radial part of the amplitude is
\begin{eqnarray}
\phi_{L}(p_0,|\textbf{p}|) = S_{++}(p_0,|\textbf{p}|) g_{L}(p_0,|\textbf{p}|),
\label{phi2g}\end{eqnarray}
with $g_{L}(p_0,|\textbf{p}|)$ being the radial part of the vertex function and
\begin{eqnarray}
S_{++}(p_0,|\textbf{p}|) = \frac{1}{(M_d/2+p_0-E_{\textbf{p}})(M_d/2-p_0-E_{\textbf{p}})},
\label{Spp}\end{eqnarray}
being the positive energy part of the propagators and the
energy $E_{\textbf{p}}=\sqrt{m^2+\textbf{p}^2}$.

Analyzing the analytic structure of expressions~(\ref{phi2g}) and~(\ref{Spp})
we can write the following expression for the poles in the $p_0$ complex plane:
\begin{itemize}

\item initial deuteron

for propagator $S_{++}(p_0,|\textbf{p}|)$:
\begin{eqnarray}
{\bar p_0}=\pm M_d/2 \mp E_{\textbf{p}} \pm i\epsilon,
\label{p0spp}
\end{eqnarray}
for functions $g_{L}(p_0,|\textbf{p}|)$:
\begin{eqnarray}
{\bar p_0}=\pm E_{\beta_k} \mp i\epsilon,
\label{p0beta}
\end{eqnarray}

\item final deuteron

for propagator $S_{++}(p_0^{\prime},|\textbf{p}^{\prime}|)$:
\begin{eqnarray}
{\bar p_0}=-(1+4\eta)M_d\pm \nonumber \\
\pm \sqrt{E_{\textbf{p}}^2+4\xi M_d|\textbf{p}|\cos{\theta}+4\xi^2 M_d^2}\mp i\epsilon,
\label{p01spp}
\end{eqnarray}

for functions $g_{L^{\prime}}(p_0^{\prime},|\textbf{p}^{\prime}|)$:
\begin{eqnarray}
{\bar p_0}=-\eta M_d\pm\nonumber\\
\sqrt{E_{\beta_k}^2+2\xi M_d|\textbf{p}|\cos{\theta}+\xi^2 M_d^2}\mp i\epsilon,
\label{p01beta}
\end{eqnarray}

\end{itemize}
with the energy $E_{\beta_k}=\sqrt{\beta_k^2+\textbf{p}^2}$, $\eta=Q^2/4M_d^2$ and
$\xi = \sqrt{\eta(1+\eta)}$.

To calculate the matrix elements~(\ref{dcur-part}) we should perform the
Wick rotation procedure. However, during the used procedure, the
poles~(\ref{p01spp}) and~(\ref{p01beta}) can get into the contour
of the $p_0$ integration. Additionally, the residue in these poles should
be taken into account. All contributions from the poles have
the threshold value on $Q^2$ which have the following form:\\

for the propagator $S_{++}(p_0^{\prime},|\textbf{p}^{\prime}|)$:
\begin{eqnarray}
Q_0^2 = M_d(2m-M_d),\nonumber
\end{eqnarray}

for the functions $g_{L^{\prime}}(p_0^{\prime},|\textbf{p}^{\prime}|)$:
\begin{eqnarray}
Q^2_k = 4M_d\beta_k.\nonumber
\end{eqnarray}

The Wick rotation procedure can be written as:
\begin{eqnarray}
i\int\limits_{-\infty}^\infty fdp_0 = \int\limits_{-\infty}^{\infty} fdp_4 \nonumber -\\
- 2\pi\sum_k{\theta (Q^2-Q_k^2)\; Res_k(f,p_0=\bar{p}_0^k)},
\label{wick}
\end{eqnarray}
where the threshold values $Q_k^2$ for the Graz II kernel are in table 1. 

\begin{table}

\begin{center}
\begin{tabular}{|c|c|}
\hline
$k$ & $Q^2_k$ (GeV/c)$^2$ \\
\hline
 0 & 0.004 \\
 1 & 1.182 \\
 2 & 1.736 \\
 3 & 3.915 \\
 4 & 5.965 \\
\hline
\end{tabular}
\end{center}

\caption {Table 1. Threshold values for the poles of the kernel}
\label{tab:1}
\end{table}

It is seen that calculations with the $Q^2 > 1.182$ (GeV/c)$^2$ must take into
account contribution from the poles of vertex function.

\section{Results and discussion}
\begin{figure}[!htpb]
  \begin{center}
    \includegraphics[width=0.5\textwidth]{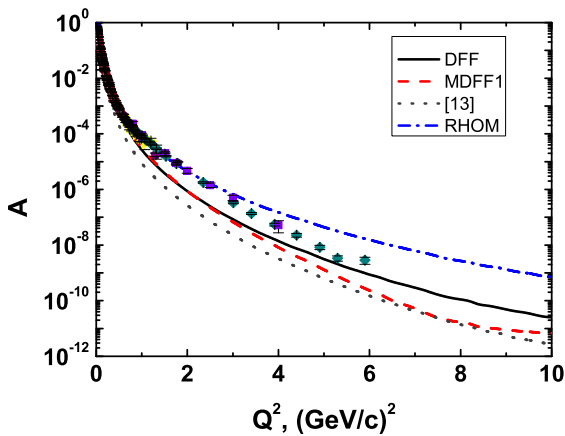}
  \end{center}
\caption{
Fig.1.
The deuteron structure function $A(q^2)$ as a function of the transfer momentum
squared. Calculations with DFF (black solid line), MDFF1 (dashed red line),
~\cite{Adamuscin:2013zoa} (gray dotted line) and RHOM (blue dashed dotted line)
nucleon form factors are shown. Experimental data are taken from~\cite{data:A}.}
\label{fig:1}
\end{figure}

\begin{figure}[!htpb]
  \begin{center}
    \includegraphics[width=0.5\textwidth]{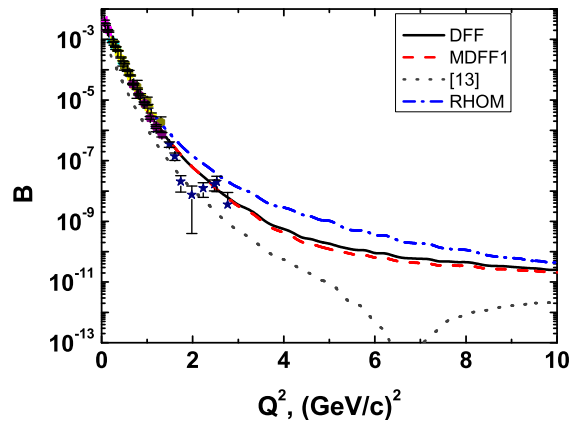}
  \end{center}
\caption{
Fig.2.
As in Fig.\ref{fig:1}, but for the structure function $B(q^2)$.
Experimental data are taken from~\cite{data:B} and~\cite{data:B1}.}
\label{fig:2}
\end{figure}

\begin{figure}[!htpb]
  \begin{center}
    \includegraphics[width=0.5\textwidth]{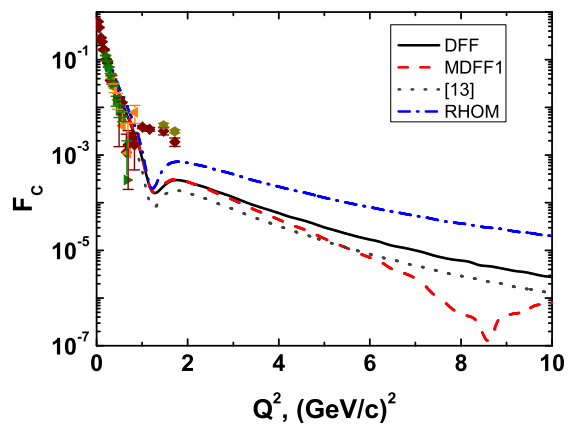}
  \end{center}
\caption{
Fig.3.
As in Fig.\ref{fig:1}, but for the charge form factor $F_{\rm C}(q^2)$.
Experimental data are taken from~\cite{data:T20}.}
\label{fig:3}
\end{figure}

\begin{figure}[!htpb]
  \begin{center}
    \includegraphics[width=0.5\textwidth]{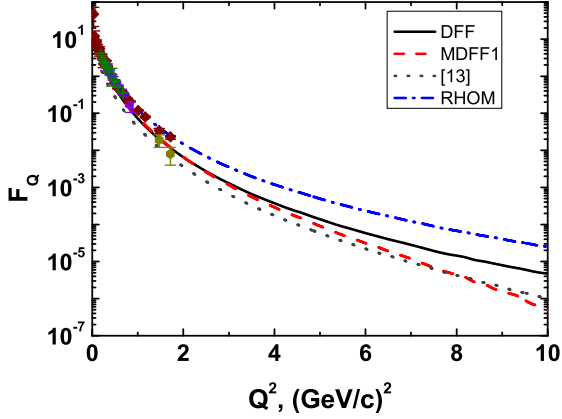}
  \end{center}
\caption{
Fig.4.
As in Fig.\ref{fig:1}, but for the quadrupole form factor $F_{\rm Q}(q^2)$.
Experimental data are taken from~\cite{data:T20}.}
\label{fig:4}
\end{figure}

\begin{figure}[!htpb]
  \begin{center}
    \includegraphics[width=0.5\textwidth]{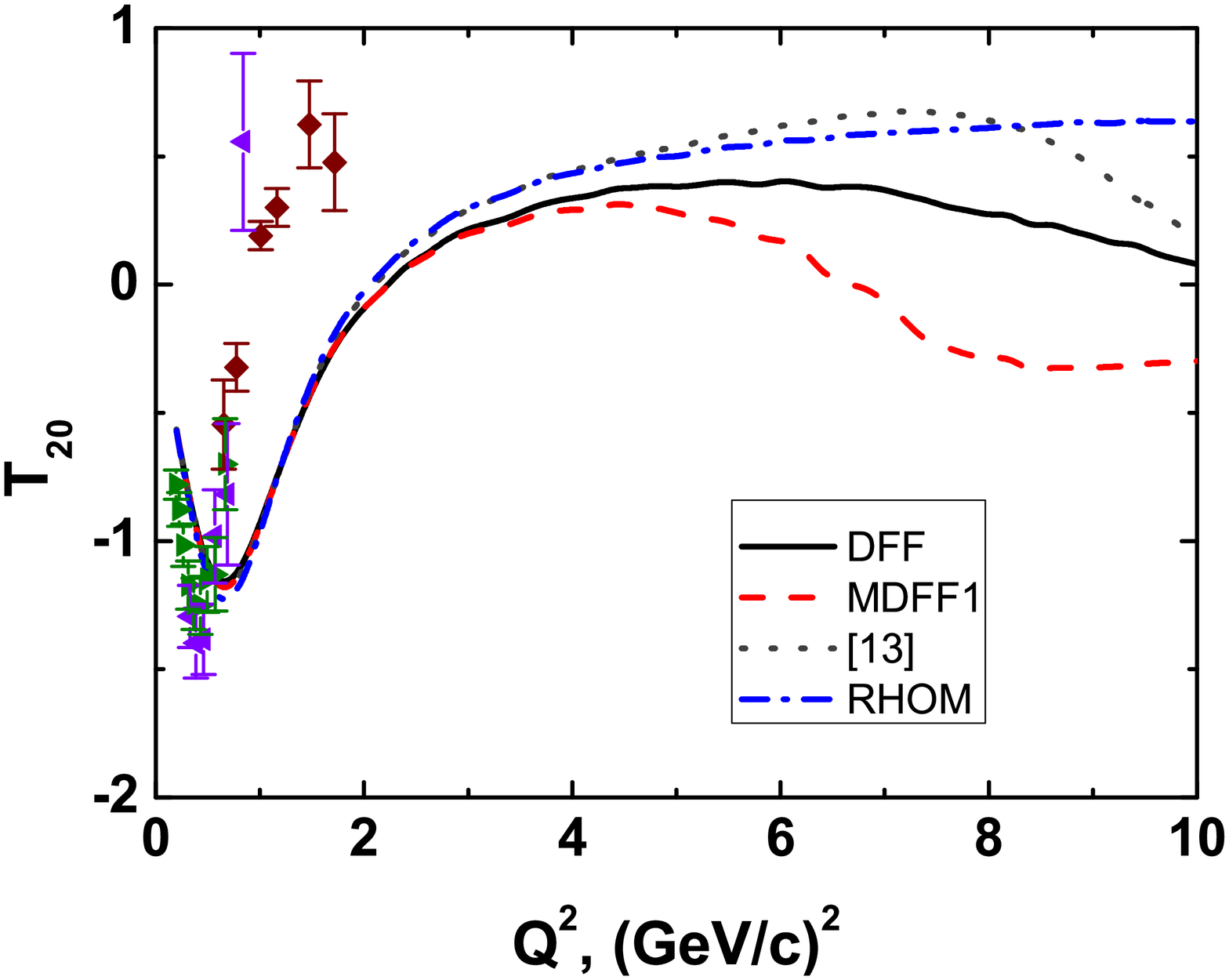}
  \end{center}
\caption{
Fig.5.
As in Fig.\ref{fig:1}, but for the tensor polarization component $T_{20}(q^2)$.
Experimental data are taken from~\cite{data:T20}.}
\label{fig:5}
\end{figure}

\begin{figure}[!htpb]
  \begin{center}
    \includegraphics[width=0.5\textwidth]{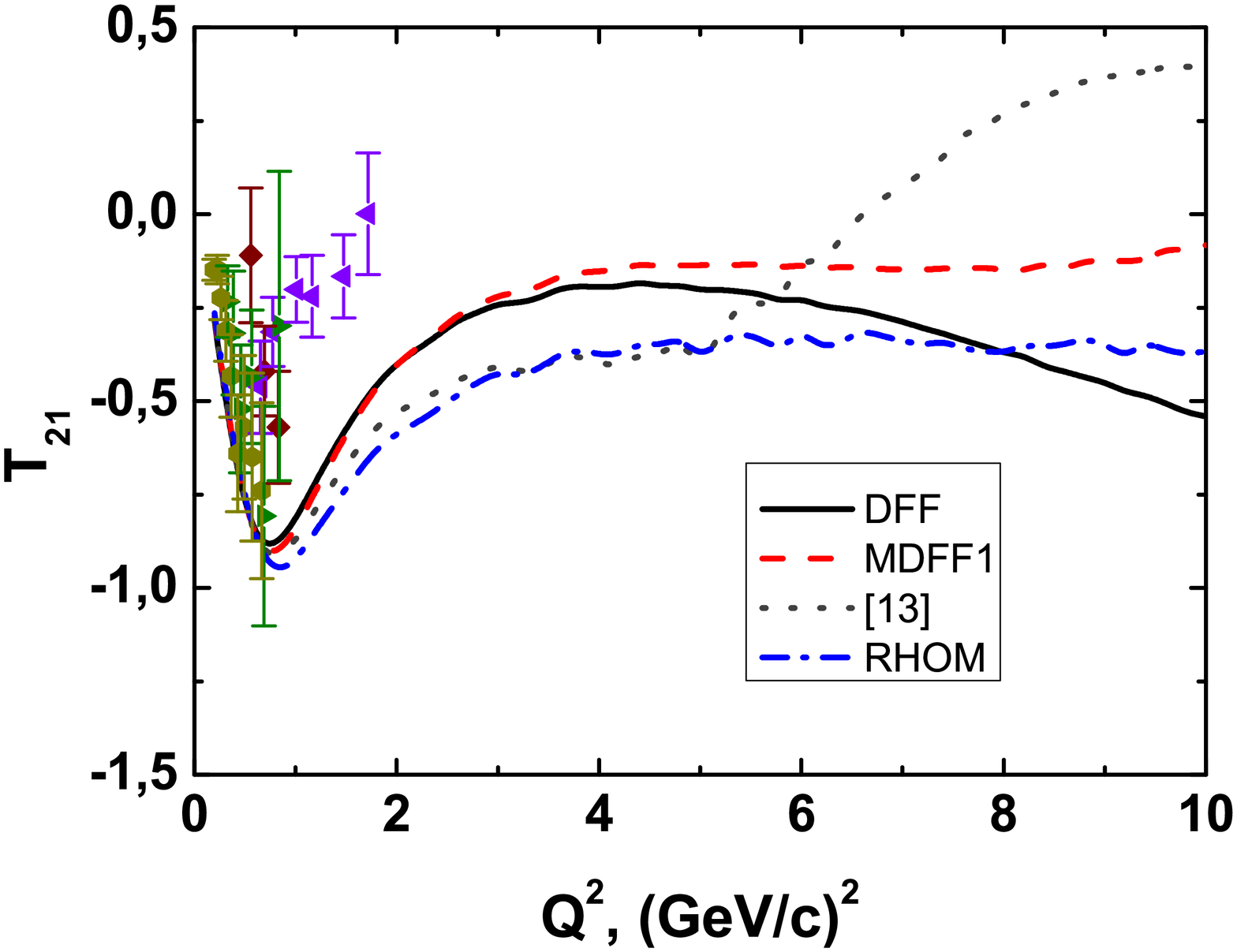}
  \end{center}
\caption{
Fig.6.
As in Fig.\ref{fig:1}, but for the tensor polarization component $T_{21}(q^2)$.
Experimental data are taken from~\cite{data:T20}.}
\label{fig:6}
\end{figure}

Figs. \ref{fig:1}-\ref{fig:6} show the influence of the considered models
of the nucleon electromagnetic form factors to the elastic electron-deuteron
scattering at high momentum transfer.

In Fig. \ref{fig:1} the deuteron structure function $A(q^2)$ is shown.
It is seen that the difference between considered models is significant and
at the $Q^2 = 10$ (GeV/c)$^2$ it reaches the value of about 2 orders for
the Ref.~\cite{Adamuscin:2013zoa} and
RHOM models. We can also see that the best model up to $Q^2 = 3$ (GeV/c)$^2$ is RHOM model
in context of experimental data coincidence, but in high-energy region the result is overestimated.
Also the results show interesting transition in MDFF1 behavior, it behaves like DFF
up to $Q^2 = 3$ (GeV/c)$^2$ and demonstrates the similar behavior with the~\cite{Adamuscin:2013zoa}
model past $Q^2 = 6$ (GeV/c)$^2$.

The deuteron structure function $B(q^2)$ is plotted in Fig.~\ref{fig:2}.
The DFF and MDFF1 results practically coincide till $Q^2 = 10$ (GeV/c)$^2$.
The RHOM and~\cite{Adamuscin:2013zoa} models are significantly differ
from previous ones and each other.
The~\cite{Adamuscin:2013zoa} have a node at approximately $Q^2=6.5$ (GeV/c)$^2$
which can be explained by the node in the proton electric
$G_{\rm E}^{\rm p}$ FF at $Q^2 \sim 11-12$ (GeV/c)$^2$.
Unfortunately there is no high-energy data for the $B(q^2)$ structure
function to make any suggestions about the physicality of such behavior.

In Figs.~\ref{fig:3} and~\ref{fig:4} the deuteron charge $F_{\rm C}(q^2)$ and
quadrupole $F_{\rm Q}(q^2)$ form factors are shown. Like for the $A(q^2)$ and $B(q^2)$
structure functions result of the RHOM model lies much higher than other results. The MDFF1 model
demonstrates the some specific feature in behavior of $F_{\rm C}(q^2)$ where
another one node appears at the $Q^2 = 8.5$ (GeV/c)$^2$ which corresponds
to the node in the proton electric $G_{\rm E}^{\rm p}$ FF. As for $F_{\rm Q}(q^2)$
MDFF1 shows the same transition like in the $A(q^2)$ structure function case.

The tensor component $T_{20}$ are shown in the Fig~\ref{fig:5}.
It is seen that all 4 models practically coincide up to $Q^2 = 4$ (GeV/c)$^2$
and~\cite{Adamuscin:2013zoa} coincides with RHOM up to $Q^2 = 8.5$ (GeV/c)$^2$.

The $T_{21}$ are shown in the Fig~\ref{fig:6}.
It is seen that all results are similar up to $Q^2 = 1$ (GeV/c)$^2$ only.
Let note that~\cite{Adamuscin:2013zoa} coincides with RHOM up to
$Q^2 = 5.5$ (GeV/c)$^2$ and DFF, MDFF1 up to $Q^2 = 4.0$ (GeV/c)$^2$

It should be noted that results for tensor components $T_{20}$ and $T_{21}$ can be
combined into two groups until the $Q^2=4-4.5$ (GeV/c)$^2$ where DFF and MDFF1
models in first pair and~\cite{Adamuscin:2013zoa} and RHOM in second pair
have a very similar behavior.
However in the region with higher energy all four models show very different trend.

\begin{figure}[!htpb]
  \begin{center}
    \includegraphics[width=0.5\textwidth]{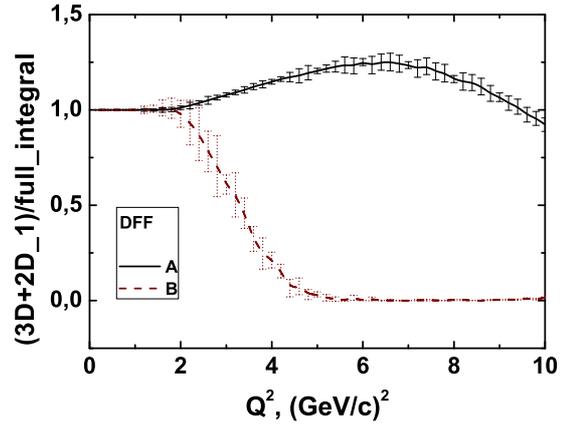}
  \end{center}
\caption{
Fig.7.
$3D+2D_1$-integral contribution to the full result.}
\label{fig:7}
\end{figure}

Fig.~\ref{fig:7} represents the influence of the vertex function poles to the
full integral value. It is seen that past $Q^2 = 2$ (GeV/c)$^2$ the poles begin
to play an important role. It is surprising that for the $B(q^2)$ structure
function at the $Q^2 = 4.5$ (GeV/c)$^2$ contribution of the poles of the separable kernel become crucial,
while for the $A(q^2)$ their contribution hardly reach $25\%$
on the whole interval. The result for the function $B(q^2)$ is because
the contributions for the $3D$- and $2D_1$-integrals (see Eq.~(\ref{wick}))
have different signs and their sum become much smaller
then contribution of the residue in the poles of deuteron vertex function.


\section{Conclusion}
In the paper the elastic electron-deuteron scattering in the relativistic
impulse approximation within the Bethe-Salpeter approach with the covariant
Graz II kernel of the nucleon-nucleon interaction is considered. Calculations
are performed at high momentum transfer up to 10 (GeV/c)$^2$. The analytic
structure of the vertex functions are taken into account. The result of
calculations with four models of the nucleon form factors are compared.
It is necessary to stress that considered high energies are required to take
into account relativistic properties of the deuteron.

As for the significant change between nucleon form factors models there is
a lack of experimental data to make any statements about which model is closer
to real physics.

The difference between presented theoretical calculations and experimental
data can be explained as it is necessary to take into account additional
contributions. Thus, the next step is to use the modern separable
kernel of nucleon-nucleon interaction~\cite{Bondarenko:2008mm}.
As an extension to the calculations, some effects should be taken into
account. Among them are the relativistic P-states in the deuteron, two-body
interaction currents and off-mass shell nucleon effects (see,
e.g.~\cite{Bondarenko:2002zz},~\cite{hamamoto:2005}).
As concerning off-mass shell effects in this approach, there is a possibility
to solve inverse task and to determine the behavior of nucleon form factors
of bound nucleon.

\section*{Acknowledgments}
Authors are grateful to the Professors S. Dubni\v{c}ka, A.Z. Dubni\v{c}kova,
Drs. C. Adamu\v{s}\v{c}\'{i}n and E. Barto\v{s} for providing data on
EM nucleon form factors (U\&A model).

\end{document}